\documentclass{llncs}

\usepackage[hide]{ed}
\usepackage[backend=bibtex,hyperref=auto,style=alphabetic,maxnames=3]{biblatex}
\bibliography{paperbib}
\usepackage{paralist}
\usepackage{graphicx}
\usepackage{wrapfig}
\author{Constantin Jucovschi}
\title{Towards an Interaction-based Integration of MKM Services into End-User Applications}
\institute{Jacobs University Bremen}

\begin{document}
\maketitle

\begin{abstract}
The Semantic Alliance (SAlly) Framework, first presented at MKM 2012, allows integration of Mathematical Knowledge Management services into typical applications and end-user workflows. From an architecture allowing invasion of spreadsheet programs, it grew into a middle-ware connecting spreadsheet, CAD, text and image processing environments with MKM services. The architecture presented in the original paper proved to be quite resilient as it is still used today with only minor changes.

This paper explores extensibility challenges we have encountered in the process of developing new services and maintaining the plugins invading end-user applications. After an analysis of the underlying problems, I present an augmented version of the SAlly architecture that addresses these issues and opens new opportunities for document type agnostic MKM services. 
\end{abstract}

\section{Introduction}
\ednote{Start that MKM is mostly done in LaTeX documents.}

% Lot of mathematical content lives in excel, cad type of applications.
A major part of digital mathematical content today is created in Computer-Aided Design (CAD) systems, spreadsheet documents, wiki pages, slide presentations, program source code. Taking advantage of Mathematical Knowledge Management (MKM) representations and services, to better manage mathematical content in these documents, is still a complex and time consuming task, often, because there is no adequate tool support. Imagine a complex CAD model composed of hundreds of components. Readers of the CAD model would benefit a lot if functional information (e.g. specifying role) could be attached to components so that: 
\begin{inparaenum}
\item [\textit{i)}] semantic services generate descriptive information about the role of the object in the model;
\item [\textit{ii)}] safety requirements matching that role could be fetched from a "safety-requirements.tex" document etc.
\end{inparaenum}
Adding functional information (e.g. encoded as a URI to an ontology concept) to CAD components can be achieved in most systems by adding a custom property to the CAD component. The services generating descriptive information and fetching safety requirements could be implemented as web services which, given the concept URI, would fetch required information. In this scenario, the engineer would rightfully consider that \textit{the MKM services are not adequately supported because he has to manually get the functional information associated to an object, change application context to a web browser and supply the functional information into the MKM service.} 
The feeling of tool support inadequacy, is not caused by the MKM services themselves, but rather by the steps (workflow) that the engineer had to perform in order to be able to consume the MKM services. I call this workflow \textit{User-Service Interaction} (USI). 

The importance of developing adequate tool support, especially for authoring, is a reoccurring topic at MKM \cite{cpoint} and also in the wider semantic technology community \cite{stenmark2004integrating, joo2011adoption, siorpaes2010human}. The arguments come from different directions. \cite{siorpaes2010human} argues that some tasks, required for semantic content creation, can be performed \textit{only} by humans; making development of computer support for tasks that allow semi-automation ever more important. \cite{stenmark2004integrating, joo2011adoption} base their intuitions on the experiences and lessons learned from deploying semantic technologies in real-world scenarios. They stress the importance of ``\textit{user-friendly modeling tools and procedures}''\cite{joo2011adoption} and share lessons learned such as ``\textit{KM systems should not be introduced as explicit stand-alone applications that user intentionally must interact with in addition to their other job responsibilities}''\cite{stenmark2004integrating}. \cite{cpoint} uses Prisoner's dilemma to explain why the long-term benefits of having semantic content fail at motivating semantic content creation. Further on, authors suggests that, improving authoring support and letting content authors reap the results of semantic content creation early on, would increase motivation for semantic content creation.

The Semantic Alliance (SAlly) Framework \cite{DavJucKoh:safusa12}, set the goal of supporting the process of building adequate tool support for Mathematical Knowledge Management services by integrating them into typical applications and end-user workflows. The framework allowed MKM services to be simultaneously integrated in several end-user applications that share the same media-type (e.g. spreadsheet documents). In this way, once an MKM service was integrated, through the SAlly Framework, with Microsoft Excel, it would work "out of the box" in Open/Libre Office Calc. Later work, added support for CAD \cite{KohlhaseEtAl:FullSemanticTransparency:2013}, text and image \cite{design2014} editors as well as allowed creation of cross-application MKM services, e.g. allowing seeing costs of a CAD component in the pricing spreadsheet document, leading to the Multi-Application Semantic Alliance Framework (MA-SAlly) \cite{KohlhaseEtAl:FullSemanticTransparency:2013}.

This paper explores extensibility challenges we have encountered in the process of integrating new MKM services into end-user applications. These challenges, described in section \ref{problem_desc}, did not depend so much on the MKM service functionality, but rather, on the type of User-Service Interaction it required. Section \ref{method}, describes how USI requirements can be decomposed into modular interfaces, that are easier to standardize but also extend. Afterwards, I present an augmented version of the Semantic Alliance Architecture that incorporates the USI modular interfaces and thus solve the extensibility problem motivating this research. The paper ends with implementation results and a conclusion.

\newcommand{\api}{\mathcal{A}}

\section{Integration Analysis of the Semantic Alliance Framework}
\label{MA-SALLY}
In this section, I want to analyze the integration strategies used in the SAlly framework \cite{DavJucKoh:safusa12} to integrate MKM services into end-user applications. I will shortly introduce each identified strategy and assess its impact by using the following cost model: 
\begin{quote}
$n \geq 2$ applications must be integrated with $m \geq 2$ MKM services. There is a cost function $C$ that computes the cost of implementing any part of the integration such that the cost of implementing any part is equal to the sum of the costs implementing its subparts. 
\end{quote}
The properties of the cost function $C$ clearly oversimplify the software development process. Yet, the extent to which these properties are used in this paper should not significantly change the validity of the computations. Also, the goal of integrating $n$ applications with $m$ services might seem unrealistic. Indeed, there are MKM services for which integration into certain applications makes little or no sense at all. On the other hand, from the experience gained with the MA-SAlly framework, MKM services such as definition lookup and semantic navigation were integrated in all invaded applications.
 
% Managing knowledge inside the same document is often at rudimentary level supported by applications.
% Managing knowledge from multiple documents and across document types is 

\begin{wrapfigure}{r}{0.5\textwidth}
\centering
\includegraphics[width=5cm]{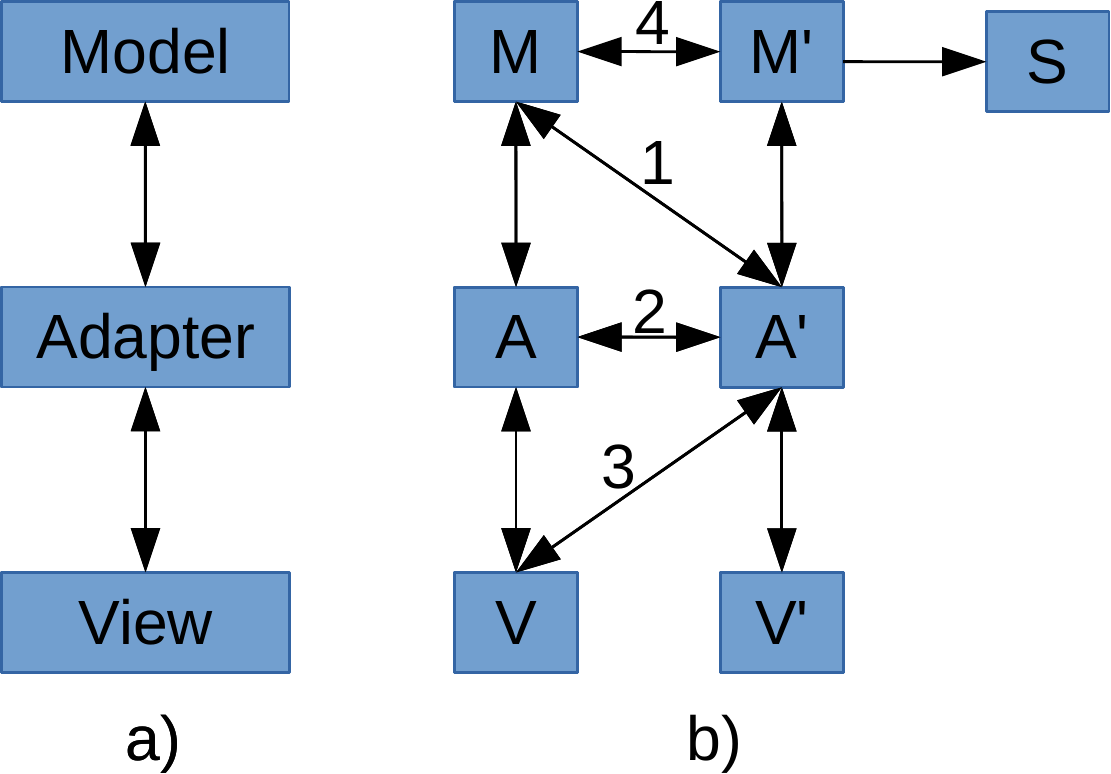}
\caption{a) the Model-View-Adapter pattern b) direct integration of a MKM service into an application. M, M' are the models, V, V' are the views and A, A' are the adapter of the application and MKM service respectively. S is an optional semantic (web-)service.}
\label{fig:mva_direct}
\end{wrapfigure}

Figure \ref{fig:mva_direct}a, shows the Model-View-Adapter (MVA) architectural pattern that, along with Model-View-Controller (MVC), are widely used in the development of applications with heavy user interaction. The MVA pattern structures 
application components into three categories: model, adapter and view. The components in the view category are responsible for the visual interface; components in the model category are responsible for application logic and components in the adapter category, mediate the interaction between view and model components. 
The MVA architectural pattern is also suitable for representing the architecture of MKM services where the model implements the MKM representation, service logic or just sends requests to an external MKM service. The view of an MKM service implements any custom dialogs, toolbars and service configuration pages. The MKM service adapter wires everything together.

MKM services that are directly integrated into an application, i.e. implemented as a plugin for that application and share the same memory space, enjoy several practical benefits. Namely, they can directly access all relevant resources the host application can provide. Figure \ref{fig:mva_direct}b, shows typical resource access patterns among end-user application components (M, A, V) and MKM service components (M', A', V'). The MKM service adapter (A'), often influences how application events are handled, how application view and model get updated (hence the edges 1, 2 and 3). The MKM model (box M') may both listen to changes and modify application model M (edge 4). When MKM services are directly integrated into an application, implementing edges 1, 2, 3 and 4 is equivalent to a performing simple API calls and hence very straightforward. A disadvantage is that directly integrating $m$ such services into $n$ applications will result in a huge cost 
\begin{equation}
\sum_{a\in App} \sum_{s \in Serv} C(M_a'^s)+C(A_a'^s)+C(V_a'^s)
\label{eq:no_integration_cost}
\end{equation}
where $M_a'^s$, $A_a'^s$ and $V_a'^s$ are the model, adapter and view that needs to be implemented to integrate MKM service $s$ into an application $a$. 

\begin{figure}
\centering
\includegraphics[width=0.8\textwidth]{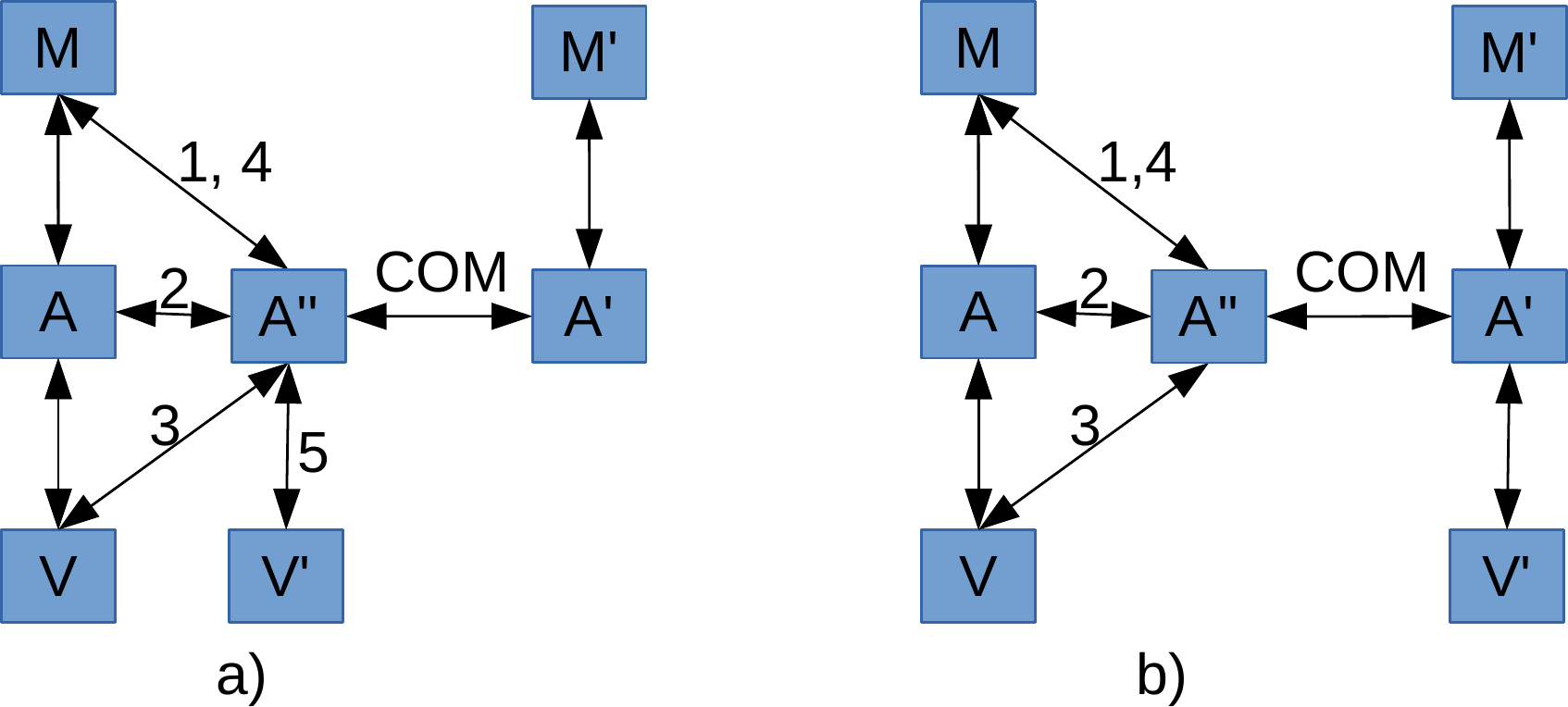}
\caption{a) refactoring MKM service model and adapter into a Web-Service b) refactoring MKM service view into the Web-Service. }
\label{fig:integration_ws}
\end{figure}

A natural way of reducing the cost in equation \ref{eq:no_integration_cost}, is to refactor MKM service models and adapters into standalone web services (Figure \ref{fig:integration_ws}a). The MKM adapter and model ($A'$, $M'$) are no longer part of the application but still need to communicate with it. This is achieved through an end-user application plugin $A''$, that allows $A'$ and $M'$ to access the same resources as before, i.e., edges $1, 2, 3$ and $4$, in Figure \ref{fig:mva_direct}b by communicating through a network channel COM. Additionally, $A''$ also needs to allow $A'$ to communicate with $V'$ (edge 5 in Figure \ref{fig:integration_ws}a). The cost of this integration strategy is:
\begin{equation}
\sum_{a\in App} \sum_{s \in Serv} C(A_a''^s)+C(V_a'^s) + \sum_{s \in Serv} C(A_{*}'^s) + C(M_{*}'^s)
\label{eq:refactor_integration_cost}
\end{equation}
where $A_{*}'^s$ and $M_{*}'^s$ are the adapter and model of MKM service $s$ implemented as a standalone service; $A_a''^s$ is the application plugin allowing standalone service $s$ to communicate with application $a$. At this point, $A_{*}'^s$ and $M_{*}'^s$ can be running on a different server, can be implemented in any language and optionally even be part of the service $S$ from Figure \ref{fig:mva_direct}b. On the other hand, implementation costs of the application plugin $A_a''^s$ can become very high as they depend on the complexity of access patterns between MKM service and application resources (i.e. edges 1, 2, 3, 4 and 5).

\begin{figure}
\centering
\includegraphics[width=0.8\textwidth]{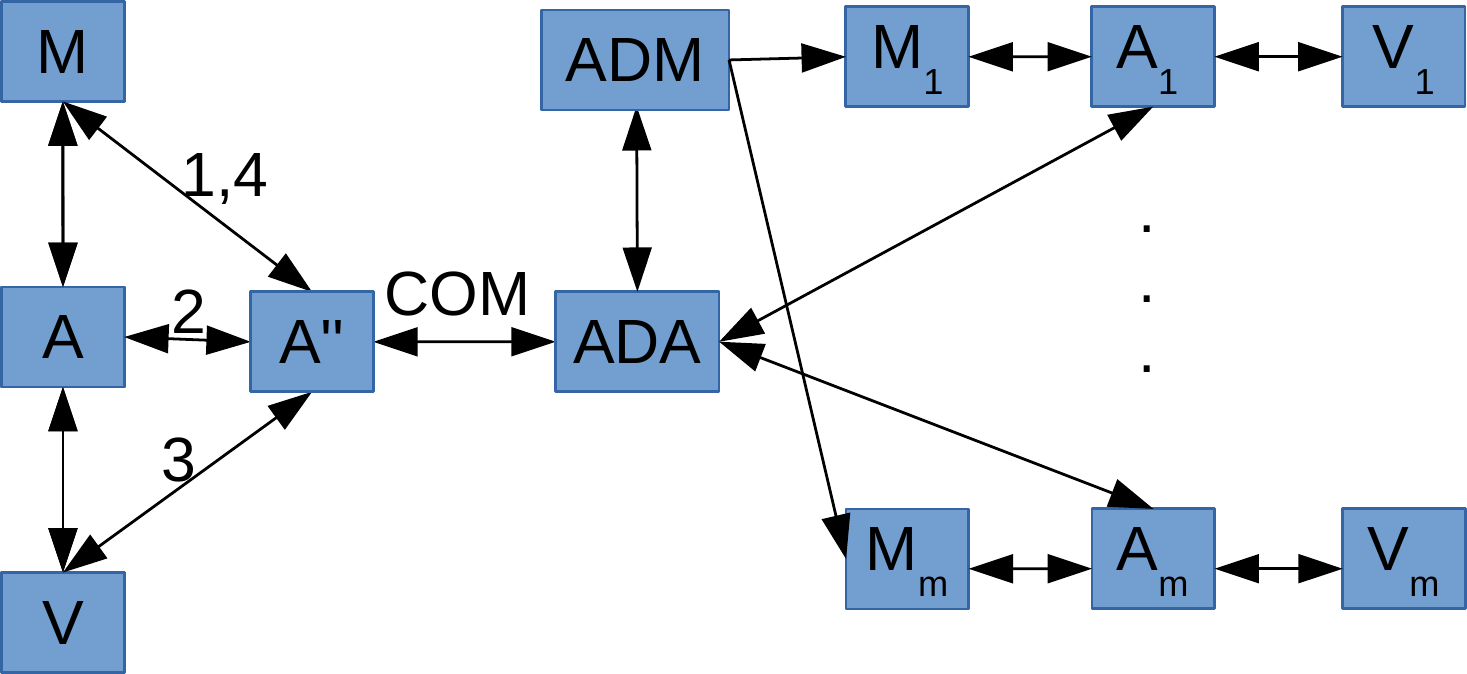}
\caption{Architecture of the Semantic Alliance Framework from an integration perspective}
\label{fig:integration_strategies}
\end{figure}

\begin{figure}
\centering
\includegraphics[width=0.7\linewidth]{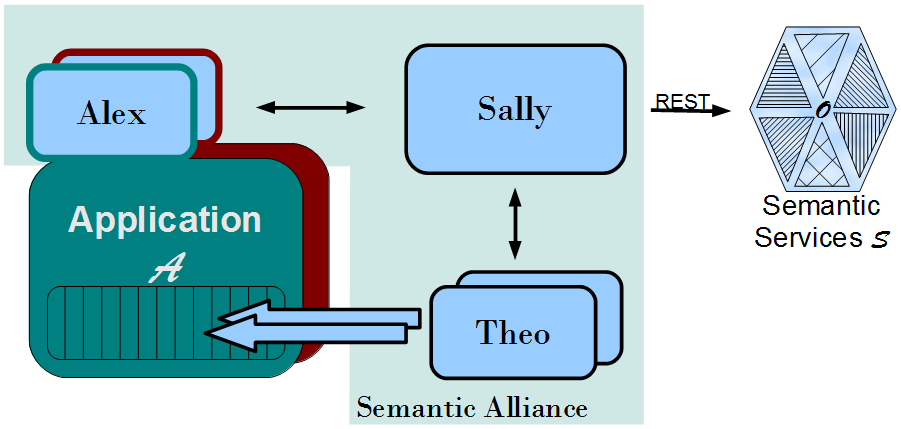}
\caption{SAlly Architecture from \cite{DavJucKoh:safusa12}}
\label{fig:sally_arch}
\end{figure}

To further reduce costs, SAlly refactors $V'$ into the same standalone service as $M'$ and $A'$ (Figure \ref{fig:integration_ws}b). Generally, this may be achieved in two ways: \begin{inparaenum}
\item [\textit{i)}] using the graphical toolkit of the language in which $V'$ is implemented --- this requires $V'$ to run on the same computer as the end-user application. 
\item [\textit{ii)}] using an external interaction interpreter (e.g. web browser) running on the same computer as the end-user application.
\end{inparaenum}
The SAlly architecture uses the second approach where a general purpose screen manager, called Theo (Figure \ref{fig:sally_arch}), runs on the same computer as the application. The cost is further reduced to
\begin{equation}
\sum_{a\in App} \sum_{s \in Serv} C(A_a''^s) + \sum_{s \in Serv} C(A_{*}'^s) + C(M_{*}'^s)+C(V_{*}'^s)
\label{eq:theo_integration_cost}
\end{equation}

Further optimizations come from standardizing and reusing plugins $A_a''^s$ for multiple services. The SAlly architecture choses to standardize $A_a''^s$ by application type (e.g. spreadsheet documents). Namely, for application type \textit{T} and a set of services $S(T)$ for that application type, one could define an API $\api(T)$ that can serve all services in $S(T)$ i.e. 
\begin{equation}
\forall a\in App(T). \forall s\in S(T). A_a''^s \subset \api(T)
\label{eq:APIT}
\end{equation}
Hence, the cost equation can be updated to
\begin{equation}
\sum_{T\in AppType}\left( \sum_{a \in App(T)} C(\api_a(T)) + \sum_{s \in S(T)} C(A_{*}'^s) + C(M_{*}'^s)+C(V_{*}'^s)\right)
\label{eq:alex_refactor}
\end{equation}
where $\api_a(T)$ is the implementation of API $\api(T)$ in application $a$ and corresponds to the application dependent invaders called Alexes in \cite{DavJucKoh:safusa12} (Figure \ref{fig:sally_arch}). 

Considering that the SAlly Framework, was specifically designed for integrating semantic services into applications, most MKM service models $M_{*}'^s$ require persistent storage for semantic information in the document. For example, most MKM services for spreadsheet documents, require functionality allowing semantic information to be attached to sheets and cell ranges. Instead of letting each MKM service model implement its own persistence strategy, an Abstract Document Model $ADM(T)$ (application-type dependent), was introduced (Figure \ref{fig:integration_strategies}), that served as a common semantic layer that all MKM services could reuse. Similarly, an Abstract Document Adapter (ADA) was implemented to ease communication and coordinate how multiple MKM services process events. The $ADM$, $ADA$, $A_{*}'^s, M_{*}'^s, V_{*}'^s$ components in Figure \ref{fig:integration_strategies} are part of the box ``Sally'' in Figure \ref{fig:sally_arch}.

\section{Problem Description}
\label{problem_desc}
%For example, a MKM service could make the spreadsheet user aware, that some values are invalid either by selecting those value or coloring them in red. To implement selection of values, a developer can use the APIs provided by the framework. Since coloring of cells is not a function Sally APIs support, a developer would have to extend and recompile Excel and LibreOffice plugins and extend SAlly communication APIs. The second option is clearly much more expensive but independent of the MKM service functionality.

The first problem of the SAlly architecture comes from equation \ref{eq:APIT}, namely, $\api(T)$ is defined as an API that can serve all services $S(T)$. Defining an $\api(T)$ for a finite set $S(T)$ is not very hard but generally, $S(T)$ is infinite. In practice, one defines an $\api(T)$ to support currently available MKM services. However, when a new MKM service needs to be integrated, which requires support to e.g. create a new sheet and the current $\api(spreadsheet)$ does not support this operation, $S$ cannot be implemented unless $\api(spreadsheet)$ is extended. Considering that multiple applications implement API $\api(T)$ (through plugin $\api_a(T)$), constantly changing $\api(T)$ becomes a bottleneck and a source of errors. Conversely, if an application can only support 95\% of $\api(T)$, it cannot be (safely) integrated even with services that do not need the rest $5\%$ of the $\api(T)$. 

Defining the abstract document model $ADM(T)$ for an application type, shares similar issues as defining $\api(T)$. This became very apparent when we tried to define an abstract document model for CAD systems \cite{KohlhaseEtAl:FullSemanticTransparency:2013}. Most CAD systems share the concept of an assembly which defines the position in space of CAD parts or nested CAD assemblies. At the assembly level, one could define an $ADM(T)$ capturing various relationships among assembly components that could be applied to most CAD systems. The interesting geometrical properties, however, are only represented in the CAD parts, which, may vastly differ even within the same CAD system (especially if CAD parts are imported from other systems). Defining an $ADM(T)$ capturing parts data is so time-consuming that it is not worth the integration benefits it brings.

Defining an application type dependent $\api(T)$ and $ADM(T)$ makes a lot of sense when the invaded applications are similar in most respects as LibreOffice Calc and Microsoft Excel (used in the original SAlly paper) are. But the list of such examples is not very long. Most applications are, to some extent, unique but still share a great deal of concepts with other applications. 

\section{Method}
\label{method}

This paper aims at reducing the dependency of integration strategies on the application type. Equation \ref{eq:theo_integration_cost} (in section \ref{MA-SALLY}) is the last integration strategy in the SAlly architecture that does not depend on the notion of application-type. At that point, the biggest cost factor is contributed by
\begin{equation}
\sum_{a\in App} \sum_{s \in Serv} C(A_a''^s)
\label{eq:cost_usi_req}
\end{equation}
where $A_a''^s$ is a plugin in application $a$ that allows MKM service $s$ to e.g. insert new data into the active document, highlight some part of the document or get notified when another object is selected and so on. \textit{$A_a''^s$ is the critical component that enables User-Service Interaction between MKM service $s$ and application $a$}. An important aspect is that, while $A_a''^s$ enables service $s$ to interact with application $a$, it does not implement the interaction itself. The implementation of the User-Service Interaction is part of the MKM service.

One of the main purposes of MKM User-Service Interactions is to support the process of aligning document content with concepts in some ontology. Typical tasks include: annotating content, making relationships explicit, validating content and managing changes. The features and the types of interactions offered by USIs strongly depend on the semantic format/ontology and often dependent only slightly on the format of the document that is semantically annotated. Consider a high-level description of the OMDoc document ontology: 
\begin{quote}
An OMDoc document organizes content into theories that relate to each other either by import or view relations. Theories may contain symbols that are transported from one theory to another using the relations among theories. 
\end{quote}
This description clearly specifies requirements regarding the types of annotations (theories and symbols) and relationships among them (theories contain symbols and relate to other theories) that OMDoc authoring services need to support. The requirements on the host document format are rather implicit: relations need to be persistent and react in sensible ways to document changes. Furthermore, validation services that make sure that imports are valid, that there are no cyclic dependencies, that all symbols are defined --- require only information regarding OMDoc relations are hence completely independent of host document format. 

From the experience gained through the Semantic Alliance Framework, I compiled a list of common and application-type independent functionality that $A_a''^s$ typically need to implement. Namely:
\begin{compactenum}
\item [\bf{content selection}] --- is by far the most common functionality USIs require. It allows the user to point MKM services to document objects, and reversely, MKM services to point the user to document objects. It is heavily used for enriching and modifying semantic content.
\item [\bf{semantic storage}] --- stores and retrieves semantic information associated to document objects. One can differentiate between document and object level semantic information.
\item [\bf{context menus}] --- provide a natural way of accessing object specific interactions. 
\item [\bf{application focus}] --- used in the multi-application context and allows changing focus to a certain application. Is usually used in combination with content selection.
\item [\bf{content marking}] --- used to visually mark (e.g. highlight) document content in a non-persistent way. Very useful for projecting multiple types of semantic content to document objects as well as selecting/deselecting multiple objects without the danger of clicking the wrong button and loosing the whole selection. 
\end{compactenum}
The intuition behind this paper, is that major parts of the User-Service Interactions that MKM services require, can be implemented on top of document-type independent interactions such as the ones above. So an annotation service for OMDoc format could be defined as follows: when the user requests the context menu for some selected object X, and there is no semantic annotation about X in the semantic storage, add the menu item ``Annotate as OMDoc module'' to the context menu, and so on. Depending on document type, selected object X can be a text fragment, a cell range, a CAD assembly. In the same time, MKM services might require very specific, application dependent information which also need to be supported e.g. position of geometrical assemblies. 

The problem of reducing the cost in equation \ref{eq:cost_usi_req}, is essentially a problem of defining reusable interfaces in a distributed setting. Any interface $X$, that can be reused by $k \geq 2$ MKM services, reduces the cost in equation \ref{eq:cost_usi_req} by $n\left(k-1\right)C(X)$ i.e. each application must implement $X$ once to then reuse it for $k$ services. One can iterate this process and a define set of reusable, modular interfaces $\mathcal{M} = \{M_1, M_2, ...\}$ such that 
\begin{equation}
A_a''^s = M_a^{s_1} \circ M_a^{s_2} \circ \ldots \circ M_a^{s_k},
\end{equation}
i.e. one can represent $A_a''^s$ as composition of several implementations of 
modular interfaces ($M^{s_1}, \ldots M^{s_k} \in \mathcal{M}$) for application $a$. Substituting this representation in equation \ref{eq:cost_usi_req} and removing duplicate module implementations for the same application, results in
\begin{equation}
 \sum_{a\in App} \sum_{m \in \mathcal{M}(S)} C(M_a^m)
 \label{eq:cost_mod}
\end{equation}
where $\mathcal{M}(S)$ is the set of all modules in $\mathcal{M}$ that are required to modularize all $A_a''^s$ in $s\in Serv$. 

The total cost in equation $\ref{eq:cost_mod}$, depends on the ability of representing the functionality, a MKM User-Service Interaction requires, in a reusable modular manner. Setting $\mathcal{M} = \left\{A''^s | s \in Serv \right\}$, i.e. no reuse possible, makes cost in equation $\ref{eq:cost_mod}$ equal to the one in equation \ref{eq:cost_usi_req}.

\section{Augmented Semantic Alliance Architecture}
\label{ammend-arch}
The Augmented Semantic Alliance Architecture assumes, that a set $\mathcal{M} = \{M_1, M_2, ...\}$  of reusable modular interfaces are defined and publicly available to the MKM community. It also assumes that the User-Service Interaction of any MKM service $s$, can be achieved by combining some set of modular interfaces that I define as $\mathcal{M}(s) \subseteq \mathcal{M}$. If for some service $s$ this is not possible, the set $\mathcal{M}$ is extended with the necessary functionality.

Each application $a$, can implement a subset of the modular interfaces denoted with $\mathcal{M}(a) \subseteq \mathcal{M}$. A MKM service $s$ can be integrated into application $a$ if $\mathcal{M}(s) \subseteq \mathcal{M}(a)$ i.e. application $a$ implements all modules required by MKM service $s$. To allow the possibility of having the equivalent of Abstract Document Models from the SAlly architecture, modular interfaces may depend on other modular interfaces (typical restrictions on circular dependencies apply). 
\begin{figure}
\centering
\includegraphics[width=0.8\linewidth]{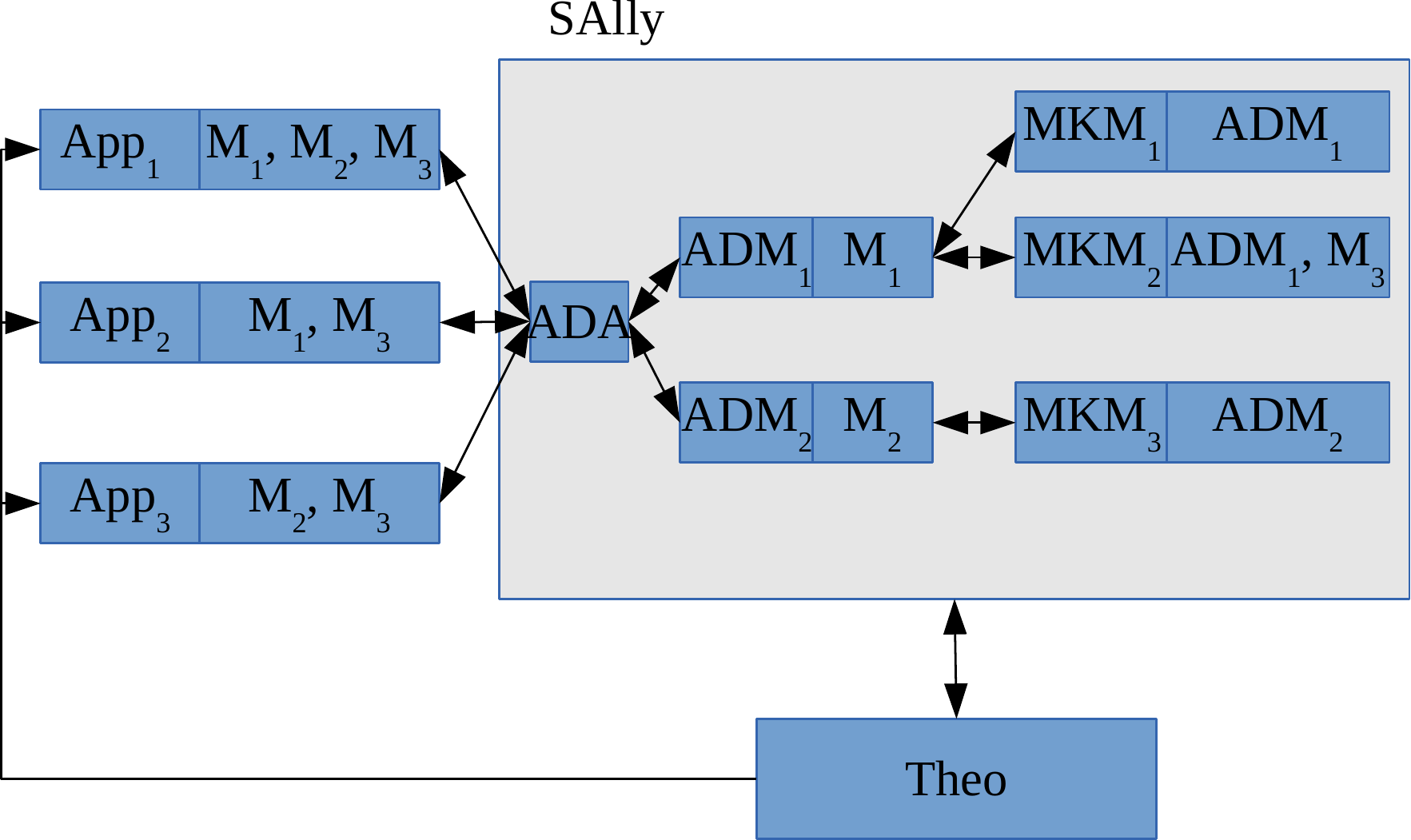}
\caption{Integration of three MKM services with three applications using the Augmented Semantic Alliance Architecture}
\label{fig:new_sally}
\end{figure}

Figure \ref{fig:new_sally}, shows an example how three applications ($App_1$, $App_2$, $App_3$) can be integrated with three services $(MKM_1, MKM_2, MKM_3)$. 
Applications $App_1$, $App_2$, $App_3$ implement modular interfaces $\left\{M_1, M_2, M_3\right\}$, $\left\{M_1, M_3\right\}$ and $\left\{M_2, M_3\right\}$ respectively. Service $MKM_1$, requires abstract document model $ADM_1$ which, in turn, requires module $M_1$ hence $\mathcal{M}(MKM_1)=\left\{M_1\right\}$. This means that one can integrate service $MKM_1$ into $App_1$ and $App_2$. Similarly, $\mathcal{M}(MKM_2)=\left\{M_1, M_3\right\}$ and so can be integrated only in $App_1$ and $\mathcal{M}(MKM_3)=\left\{M_2\right\}$ and can be integrated into $App_1$ and $App_3$.

From this example, one can see that the reuse strategy the Augmented Semantic Alliance Framework has, is more flexible than the one presented in \cite{DavJucKoh:safusa12}. Namely, $App_1$, $App_2$ and $App_3$ clearly share common concepts, but assigning them an application type that guides reuse strategy, reduces reuse opportunities. Also, the augmented Sally architecture solves the extensibility problems described in section \ref{problem_desc} as reuse is no longer associated with application type. Additionally, the augmented architecture also allows abstract document models to be implemented in the end-user applications themselves, if that makes integration easier. 

\section{Implementation}
\label{impl}
To test the interaction based method of integrating MKM services into applications, I chose three simple MKM services that cover four, out of five types of application independent interactions presented in section \ref{method}. Namely: content selection, semantic storage, context menu and application focus. These services were integrated into five applications that are part of the LibreOffice suite: Writer (rich text processor), Calc (spreadsheet application), Impress (slide presentations), Draw (graphic documents) and Base (database manager). In this section, I want to shortly introduce the MKM services that I integrated into the LibreOffice suite. These services are quite different and yet have almost identical User-Service Interaction requirements. The section will end in a discussion about the effort it took to perform this integration.

\begin{figure}
\centering
\includegraphics[width=0.7\linewidth]{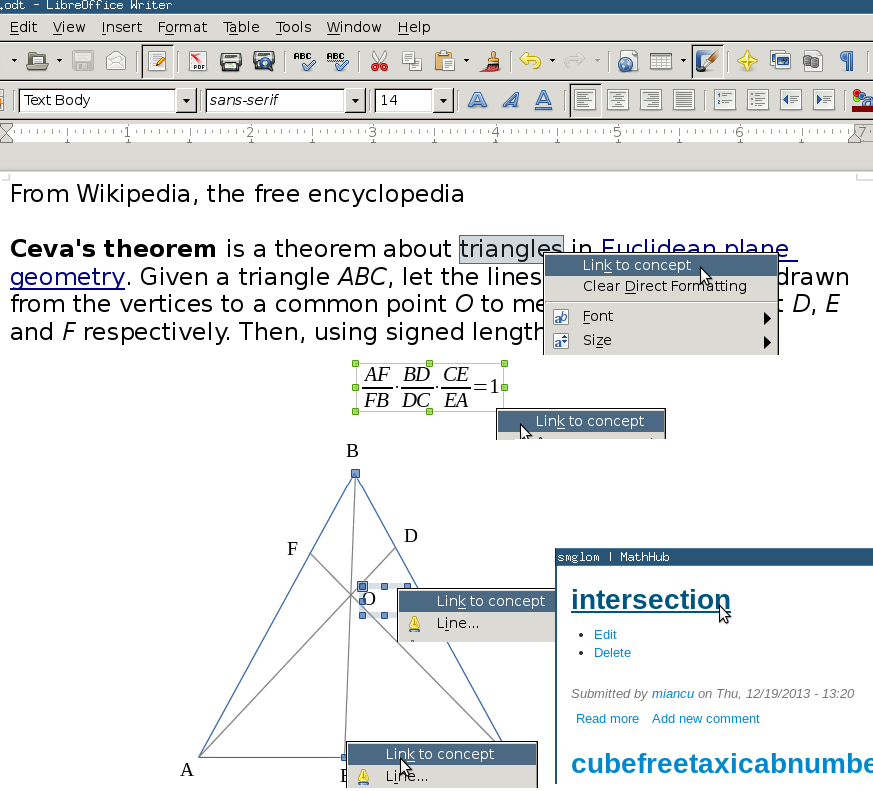}
\caption{Mash-up of several screenshots demonstrating the types of LibreOffice Writer objects that can be connected to ontology concepts by the Concept Linker.}
\label{fig:concept_linker}
\end{figure}

The \textbf{Concept Linker} service, allows linking document contents to ontology concepts stored in the \url{MathHub.info}\cite{IanJucKoh:sdm14} portal. The service adds a new "Link to concept" item in the context menu of the application, if currently selected object(s) can store semantic information. After clicking on the "Link to concept" context item, a window is generated by the Theo screen manager (for more information see \cite{DavJucKoh:safusa12}) where one can choose the concept to which selected objects should be linked to. Figure \ref{fig:concept_linker} depicts the concept linker service in a LibreOffice Writer document. One can see that text fragments, formulas, drawings (segment BE) and text boxes can be linked to ontology concepts. Also, one can see the Theo window allowing the user to link text box ``O'' to a concept from the \url{MathHub.info} portal.

\textbf{Definition lookup} service retrieves the definition of the ontology concept associated to the currently selected object. The service adds a new "Get definition" item in the context menu of the application, if currently selected object has a concept linked to it. The definition associated to the concept is displayed in a window generated by the Theo screen manager. 

\begin{figure}
\centering
\includegraphics[width=0.8\linewidth]{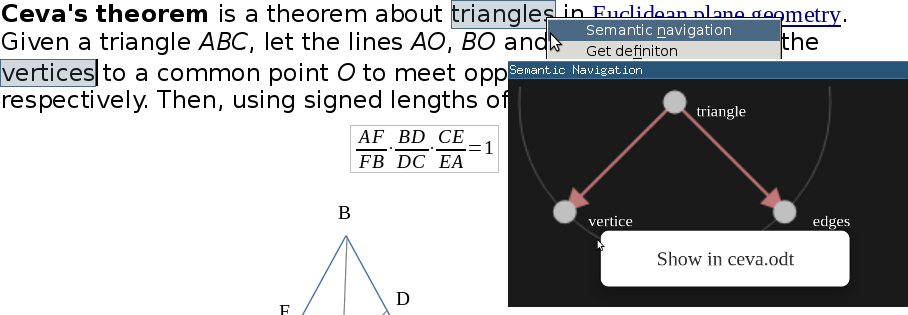}
\caption{Mash-up of several screenshots demonstrating Semantic Navigation service. }
\label{fig:sem_nav}
\end{figure}

\textbf{Semantic Navigation} presents the user with a graphical representation of ontology relations associated to the concept selected in the document. Just like the concept linker and definition lookup services, it adds a context menu item ``Semantic navigation'' which triggers creation of a new window presenting the ontology relations (Figure \ref{fig:sem_nav}). Additionally, if the user right-clicks on a node of a related concept e.g. vertice, and there is an object \textit{V} in the document linked to the concept of vertice, the user is given the possibility to change the document focus to object \textit{V}. 

The following table summarizes the types of objects that can be annotated and support definition lookup and semantic navigation services.
\begin{center}
\begin{tabular}{|c|p{9cm}|}
\hline Application & Types of document object\\ 
\hline All applications & images, math formulas, text boxes and shapes \\
\hline Writer & text fragments \\ 
\hline Calc & cell ranges, charts \\ 
\hline Impress & slides, text fragments \\ 
\hline Draw & text fragments \\
\hline Base & tables, forms, queries and reports\\
\hline 
\end{tabular} 
\end{center} 

Even though the document models of the applications in the LibreOffice suite are very different, the LibreOffice API provides several document type independent mechanisms to access/modify document's content and meta-data. The implementation of the content selection, context menu, semantic storage and application focus  interactions required by the three MKM services, were implemented using the document model independent API provided by LibreOffice. This meant that only one LibreOffice plugin had to be implemented which required a similar amount of effort as creating one Alex invader in the SAlly architecture ($\approx$ 1 week). Integrating the same services into the SAlly architecture from \cite{DavJucKoh:safusa12}, would have required a 5 weeks investment into the Alex (invader) plugins. 

%These are all fairly different document models. The reason why creating a definition lookup service for all these types of documents was possible is that LibreOffice provides document model independent interfaces through which many of the elements inside the document model can be accessed. For the definition lookup services I used three interfaces:
%\begin{compactenum}
%\item XSelectionChangeListener --- event listener that gives access to the objects selected by the user. 
%\item XIndexAccess enables enumeration of selected objects, in case multiple objects were selected.
%\item XPropertySet --- a collection of properties (e.g. font, width, color) associated to an object. Most objects have a property called "UserDefinedAttributes" that allows associating custom information to LibreOffice objects.
%\end{compactenum}
%Using these three generic interfaces, one can implement document model agnostic interactions, allowing the user to select one or more objects and associate a definition to them. It is worth mentioning that LibreOffice treats the custom properties as it would treat the e.g. bold face property i.e. it can be copied and a character inserted inside a bold word becomes bold as well. 

%Having such a semantic friendly infrastructure in place, one can create a document model independent interface exposing only semantic relevant parts in a RDF-like manner. On top of this interface, one can implement both generic RDF and ontology specific interactions.

\section{Conclusion}
This paper tackles the problem of MKM services lacking adequate tool support when integrated into end-user applications. The Semantic Alliance framework \cite{DavJucKoh:safusa12}, developed for reducing MKM service integration costs without sacrificing usability, was successfully used for integrating MKM services into spreadsheet\cite{DavJucKoh:safusa12}, CAD\cite{KohlhaseEtAl:FullSemanticTransparency:2013}, text and image \cite{design2014} processing software. In the process of developing these integrations, several extensibility problems of the framework became very apparent and yet could not be predicted from the original architecture presented in \cite{DavJucKoh:safusa12}.

First contribution of this paper is the in-depth integration analysis that helped identifying the reasons for the extensibility challenges in the SAlly framework. In particular, this analysis captures and categorizes the hidden costs associated with decoupling a MKM service into a standalone entity (cost of implementing $A_a''^s$ in section \ref{MA-SALLY}). While the integration analysis is strictly used for the SAlly framework, it can be reused for other integration efforts in MKM such as integration of theorem provers into authoring solutions. 

The second and main contribution of this paper is the Augmented Semantic Alliance architecture that solves the extensibility problems presented in section \ref{problem_desc} and enables more flexible reuse strategies. First experiments with the new architecture show that end-user applications, sometimes provide major reuse opportunities and that the new architecture can take full advantage of them.

\printbibliography

\end{document}